\documentclass[useAMS,usenatbib]{mn2e}
\usepackage{natbib}
\usepackage{epsfig}
\usepackage{amsbsy}
\usepackage{hyperref}
\usepackage{ amssymb }

\providecommand{\adsurl}[1]{\href{#1}{ADS}}

\newcommand{\eg}{{e.g.}}
\newcommand{\lsim}{\,\lower2truept\hbox{${<\atop\hbox{\raise4truept\hbox{$\sim$}}}$}\,}
\newcommand{\gsim}{\,\lower2truept\hbox{${>\atop\hbox{\raise4truept\hbox{$\sim$}}}$}\,}

\newcommand{\vv}{~~~,}

\def\eg{{\rm e.g.$\,$}}

\newcommand{\be}{\begin{equation}}
\newcommand{\ee}{\end{equation}}
\newcommand{\bea}{\begin{eqnarray}}
\newcommand{\eea}{\end{eqnarray}}
\newcommand{\beann}{\begin{eqnarray*}}
\newcommand{\eeann}{\end{eqnarray*}}
\newcommand{\benn}{\begin{equation*}}
\newcommand{\eenn}{\end{equation*}}

\DeclareMathAlphabet{\mathsc}{OT1}{cmr}{m}{sc}
\def\testbx{bx}%
\DeclareRobustCommand{\ion}[2]{%
\relax\ifmmode
\ifx\testbx\f@series
{\mathbf{#1\,\mathsc{#2}}}\else
{\mathrm{#1\,\mathsc{#2}}}\fi
\else\textup{#1\,{\mdseries\textsc{#2}}}%
\fi}
\title[High-z massive clusters as a test for dynamical coupled dark energy]
{High-z massive clusters as a test for dynamical coupled dark energy}

\author[Marco Baldi \& Valeria Pettorino] 
{Marco Baldi$^{1,2}$ \&
Valeria Pettorino$^{3}$
\\
$^1$ Excellence Cluster Universe, Boltzmannstr. 2, D-85748 Garching, Germany\\
$^2$ University Observatory, Ludwig-Maximillians University Munich, Scheinerstr. 1, D-81679 Munich, Germany\\
$^3$ SISSA, Via Bonomea 265, 34136 Trieste, Italy\\
\\}

\begin{document}
\maketitle
\begin{abstract}
The recent detection by \citet{Jee_etal_2009} of the massive cluster XMMU J2235.3-2557 at a redshift $z \approx 1.4$, with an estimated mass
$M_{324} = (6.4 \pm 1.2) \times 10^{14} M_{\odot}$, has been claimed to be a possible challenge to the standard $\Lambda $CDM cosmological
model. More specifically, the probability to detect such a cluster has been estimated to be $\sim 0.005$ if a $\Lambda$CDM model with 
gaussian initial conditions is assumed, resulting in a $3 \sigma$ discrepancy from the standard cosmological model.
In this paper we propose to use high redshift clusters as the one detected in \cite{Jee_etal_2009} to compare the cosmological constant 
scenario with interacting dark energy models. We show that coupled dark energy models, where an interaction is present between 
dark energy and cold dark matter, 
can significantly enhance the probability to observe very massive clusters at high redshift.
\end{abstract}


\section{Introduction}
One of the main challenges of present cosmology is to use observations to distinguish between a cosmological constant scenario and dynamical dark energy.
More generally, it is a crucial task to devise new observational tests capable of detecting possible failures of the standard $\Lambda$CDM model, also in view of future observations, and to constrain the parameter space of alternative scenarios. 
The existence of massive clusters at high redshift has been recently used to test $\Lambda$CDM cosmologies. In particular,
weak lensing observations in a survey of $11~deg^2$ \citep{Jee_etal_2009,Rosati_etal_2009} have recently detected a cluster of mass $M_{324} = (6.4 \pm 1.2) \times 10^{14} M_{\odot}$ ($M_{200} = (7.3 \pm 1.3) \times 10^{14} M_{\odot}$) at a redshift $z \approx 1.4$. The observation of such a massive cluster at high redshift represents an extremely rare event in the context of hierarchical structure formation. 
More specifically, the probability to detect such a cluster within a $\Lambda$CDM model with gaussian initial conditions in the volume considered by the survey presented in \cite{Jee_etal_2009} and \citet{Rosati_etal_2009} has been estimated by the same authors to be $\sim 0.005$, resulting in a $3 \sigma$ discrepancy with the assumed model.
Naturally, this discrepancy can indicate that either initial conditions are not perfectly gaussian, or that the $\Lambda$CDM model itself has
to be modified. Some analysis have already been carried out in order to test the former possibility and the detection of high-z massive clusters has been 
used to constrain the level of primordial non-gaussianity \citep{Jimenez_verde_2009, Holz_perlmutter_2010, Cayon_etal_2010, Grossi_etal_2007, Sartoris_etal_2010,Hoyle_etal_2010}. In this paper, instead, 
we investigate the second option: we point out that the presence of high-z massive clusters can be used to test the cosmological constant
scenario as compared to dynamical coupled dark energy models. We therefore keep gaussian initial conditions and instead allow for the possibility that
dark energy is not a cosmological constant but rather a dynamical scalar field that can interact with cold dark matter (CDM).

We carry out N-Body simulations for a set of models of interacting dark energy with either constant or time dependent couplings.
We then compute the cumulative halo mass functions of the structures that form within these theories at different redshifts, and compare them with the outcomes of the standard $\Lambda $CDM model.
In this way we are able to show that the probability of finding massive clusters at high redshifts is significantly enhanced in 
coupled dark energy cosmologies with respect to $\Lambda$CDM.

\section{coupled dark energy}

We consider coupled dark energy models where the acceleration 
of the Universe is driven by a scalar field $\phi $ \citep{Wetterich_1988, Ratra_Peebles_1988} interacting with the CDM fluid according to the equations:
\bea
\label{cons_c} \rho_{c}' +3 {\cal H} \rho_{c} &=& - \beta \phi' \rho_{c} \\
\label{cons_phi} \rho_{\phi}' +3 {\cal H} \rho_{\phi} &=& + \beta \phi' \rho_{c} \vv 
\eea 
with $\rho_c$ the energy density of CDM and where a prime denotes a derivative with respect to conformal time. The dimensionless function 
\be \label{beta_phi} \beta \equiv - \frac{d \ln{m_c}}{d \phi} \vv \ee
where $m_{c}$ is the mass of a CDM particle, fully specifies the interaction. Note that we define densities as $\rho \equiv 8 \pi G \rho$ and that the scalar field $\phi$ 
is normalized in units of the reduced Planck mass $M = (8 \pi G)^{-1/2}$.
The scalar field energy density is $\rho_{\phi} = \frac{\phi'^2}{2 a^2} + U(\phi)$
and we consider two choices of potentials: 
\bea 
\label{U_def_ipl} U(\phi) &=& U_0 {\phi^\alpha} \\
\label{U_def_exp} U(\phi) &=& U_0 e^{- \alpha {\phi}} \eea
where $U_0$ and $\alpha$ are constants. In the present work we limit our investigation to the models specified in Table~\ref{Simulations_Table}: one model with a potential given by (\ref{U_def_ipl}) and with a constant coupling $\beta = 0.2$, labelled as ``RP'', and two models with an exponential potential (\ref{U_def_exp}) with constant coupling $\beta = 0.2$ and variable coupling $\beta (\phi ) = 0.4\cdot e^{2\phi }$, labelled ``EXP005" and ``EXP010a2", respectively \citep[see][for a detailed discussion on variable coupling models]{Baldi_2010}. 
\begin{table*}
\begin{tabular}{lcccccccc}
\hline
\begin{minipage}{60pt} Model \end{minipage} & $ U(\phi) $ & $ \alpha $ &  $\beta $ & \begin{minipage}{60pt} Box Size \\ ($h^{-1}$ Mpc) \end{minipage}
 & \begin{minipage}{60pt} Number \\ of particles \end{minipage} & \begin{minipage}{60pt} $M_{b}$ \\ ($h^{-1} M_{\odot }$) \end{minipage} & \begin{minipage}{60pt}$M_{\rm CDM}$ \\($h^{-1} M_{\odot }$) \end{minipage} & \begin{minipage}{60pt} $\epsilon _{s}$ \\($h^{-1}$ kpc) \end{minipage} \\
\hline
$\Lambda$CDM (low)& -- & -- & 0 & 320 & 256$^{3}$ & -- & 1.47 $\times $ 10$^{11} $& 25.0\\
$\Lambda$CDM (high)& -- & -- & 0 & 80 & 2 $\times $ 512$^{3}$ & 4.7 $\times $ 10$^{7}$ & 2.3 $\times $ 10$^{8}$& 3.5\\
RP & $\phi ^{-\alpha }$ & 0.143 & 0.2 & 320 & 2 $\times $ 512$^{3}$ & 4.7 $\times $ 10$^{7}$ & 2.3 $\times $ 10$^{8}$& 3.5\\
EXP005 & $e^{-\alpha \phi }$ & 0.1 & 0.2 & 80 & 256$^{3}$ & -- & 1.47 $\times $ 10$^{11}$& 25.0\\
EXP010a2 & $e^{-\alpha \phi }$ & 0.1 & $0.4\cdot e^{2\phi }$ & 320 & 256$^{3}$ & -- & 1.47 $\times $ 10$^{11} $& 25.0\\
\hline 
\end{tabular}
\caption{List of the different simulations discussed in the present work, performed with our modified version
  of {\small GADGET-2}. The low resolution simulations include only CDM particles while the two high resolution runs include uncoupled baryons with hydrodynamical forces. The last column of the table reports the gravitational softening $\epsilon _{s}$ used in each simulation.}
\label{Simulations_Table}
\end{table*}

Coupled dark energy models have been widely studied in the literature \citep[][and references therein]{Wetterich_1995, Amendola_2000, Mangano_etal_2003, pettorino_baccigalupi_2008}. 
The interaction imprints specific features on the growth of cosmic structures, as it has been shown 
both within spherical collapse models \citep{Wintergerst_pettorino_2010, Mainini_bonometto_2006} and 
at the nonlinear level by means of N-body simulations for constant \citep{Baldi_etal_2010, Maccio_etal_2004} and 
variable couplings $\beta (\phi )$ \citep{Baldi_2010}.
In particular, the interaction determines an enhancement of structure 
formation due to the presence of a long range fifth-force acting between coupled massive particles. 
In addition to this effect, conservation of momentum turns into an extra acceleration directly proportional 
to the peculiar velocity of coupled particles; at the nonlinear level, it has been shown that this velocity dependent term can
determine a reduction of the concentration of halos if the coupling function $\beta (\phi)$ is positive \citep[see again][]{Baldi_etal_2010, Baldi_2010}. 
These effects are encoded in the modified newtonian acceleration equation for coupled dark matter particles:
\begin{equation}
\label{gadget-acceleration-equation}
\dot{\vec{p}}_{i} = \frac{1}{a}\left[ \beta \frac{\dot{\phi }}{M} a \vec{p}_{i} + \sum_{j \ne i}\frac{\tilde{G}_{ij}m_{j}\vec{x}_{ij}}{|\vec{x}_{ij}|^{3}} \right] \,,
\end{equation}
where $i$ and $j$ are indices that span over all the particles of the
simulation, $\vec{p}$ is the momentum of the $i$-th particle,
and $\tilde{G}_{ij}$ is the effective gravitational constant between the $i$-th
and the $j$-th coupled particles
\be
\label{G_eff} 
\tilde{G}_{ij} = G_{N}[1+2\beta^2(\phi)], \ee where $G_{N}$ is the usual
Newtonian value.

\section{simulations}
In this Letter, we estimate how the probability of finding very massive clusters comparable
 to the XMMU J2235.3-2557 detection by \citet{Jee_etal_2009} and \citet{Rosati_etal_2009} at $z\sim 1.4$ is modified in the presence of
 a coupling between dark energy and CDM as compared to the standard $\Lambda $CDM case.
 With this aim, we use the modified version of the cosmological N-body code
 {\small GADGET-2} \citep{gadget-2} presented in 
\citet{Baldi_etal_2010} and \citet{Baldi_2010}, to which we refer for further details, to investigate nonlinear structure formation within the models described in Table~\ref{Simulations_Table}. 
We make use of the following sets of simulations:
\begin{itemize}
 \item two of the high-resolution hydrodynamical
 simulations presented in \cite{Baldi_etal_2010} for the case of a scalar field with an inverse-power law self interaction
 potential with a constant coupling $\beta = 0.2$ and for a reference $\Lambda $CDM model, both with cosmological parameters in accordance with WMAP5 results \citep{wmap5};
\item three new simulations at lower resolution in a larger cosmological box for the case of an exponential
self interaction potential with either constant ($\beta = 0.2$) or variable ($\beta =0.4\cdot e^{2\phi }$) coupling and for a reference $\Lambda $CDM cosmology, all with cosmological parameters in accordance with WMAP7 results \citep{wmap7}. 
\end{itemize}
The former set of simulations includes the cosmological fraction of uncoupled baryons on which hydrodynamical forces are computed
 with the {\em Smoothed Particle Hydrodynamics} \citep{Springel_Hernquist_2002,gadget-2} algorithm, while the latter ones are pure CDM simulations.

The two sets of simulations start at $z=60$ with exactly the same initial conditions for each set of runs. This is a conservative setup, 
since the effects of enhanced growth that we are investigating would be more pronounced if a normalization at decoupling
 ($z\sim 1100$) had been adopted. We remark that this normalization is different from the one
 adopted in most of the simulations discussed in \citet{Baldi_etal_2010} and \citet{Baldi_2010}: in these works the amplitude of linear density
 fluctuations was normalized in order to have the same $\sigma _{8}$ at the present time. In the present study, instead, we choose
 all the models with the same initial normalization of the power spectrum, which will necessarily result in different values of $\sigma _{8}$ at $z=0$ for the different cosmological models. It is important to notice that all our models start at high redshift with a normalization of the Power Spectrum which is in full accordance with the latest WMAP7 determination of the scalar perturbations amplitude from CMB data alone \citep{wmap7}.

Given this setup, we study the halo mass function at different redshifts for the groups identified in each simulation with a Friends-of-Friends (FoF) algorithm with a linking length $\lambda = 0.2 \times \bar{d}$, where $\bar{d}$ is the mean particle spacing.
We show that the interaction between dark energy and CDM results in a significantly larger number of massive halos 
at any epoch, with respect to $\Lambda$CDM.

\section{Results}

\begin{figure*}
\includegraphics[scale=0.45]{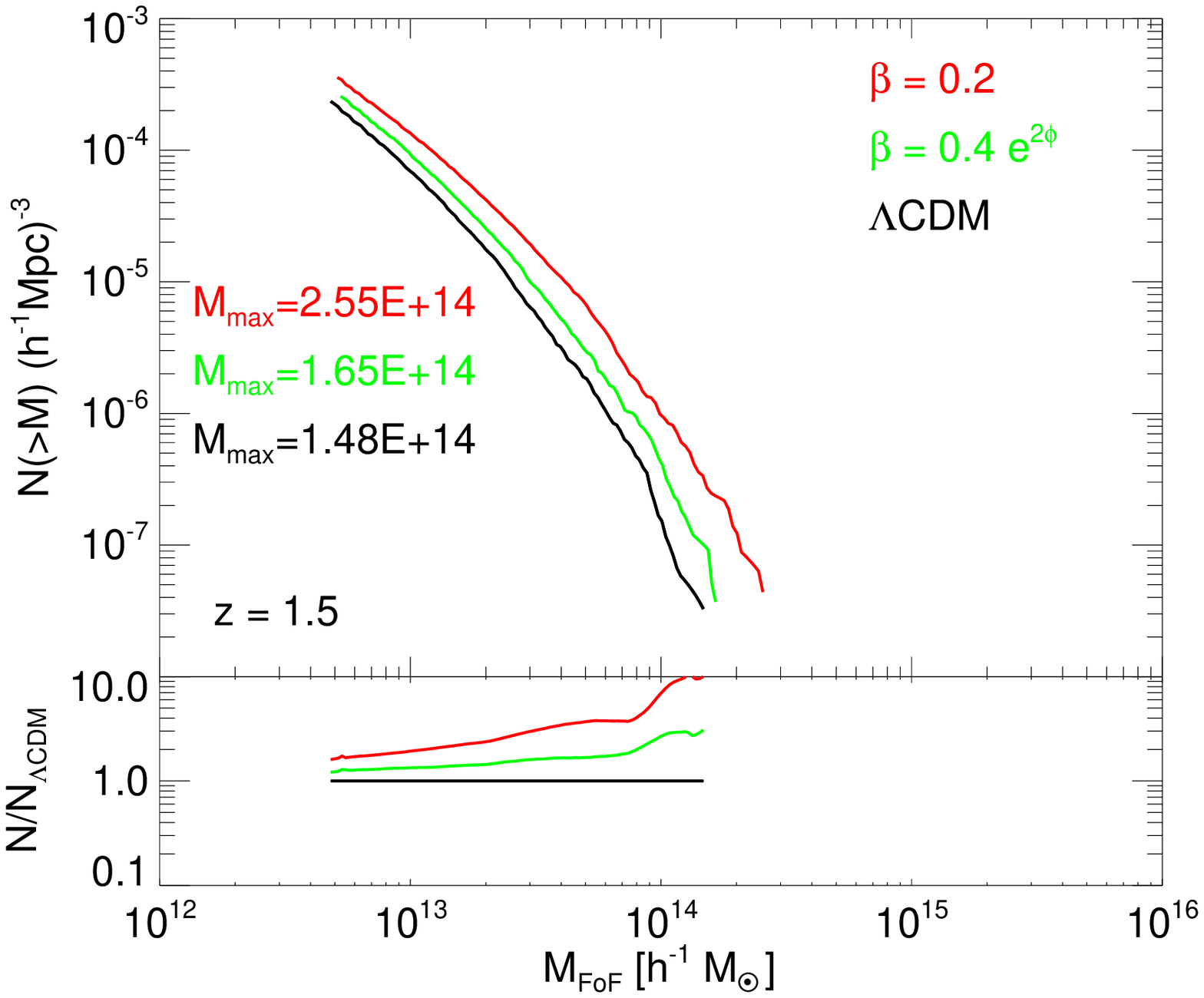}
\includegraphics[scale=0.45]{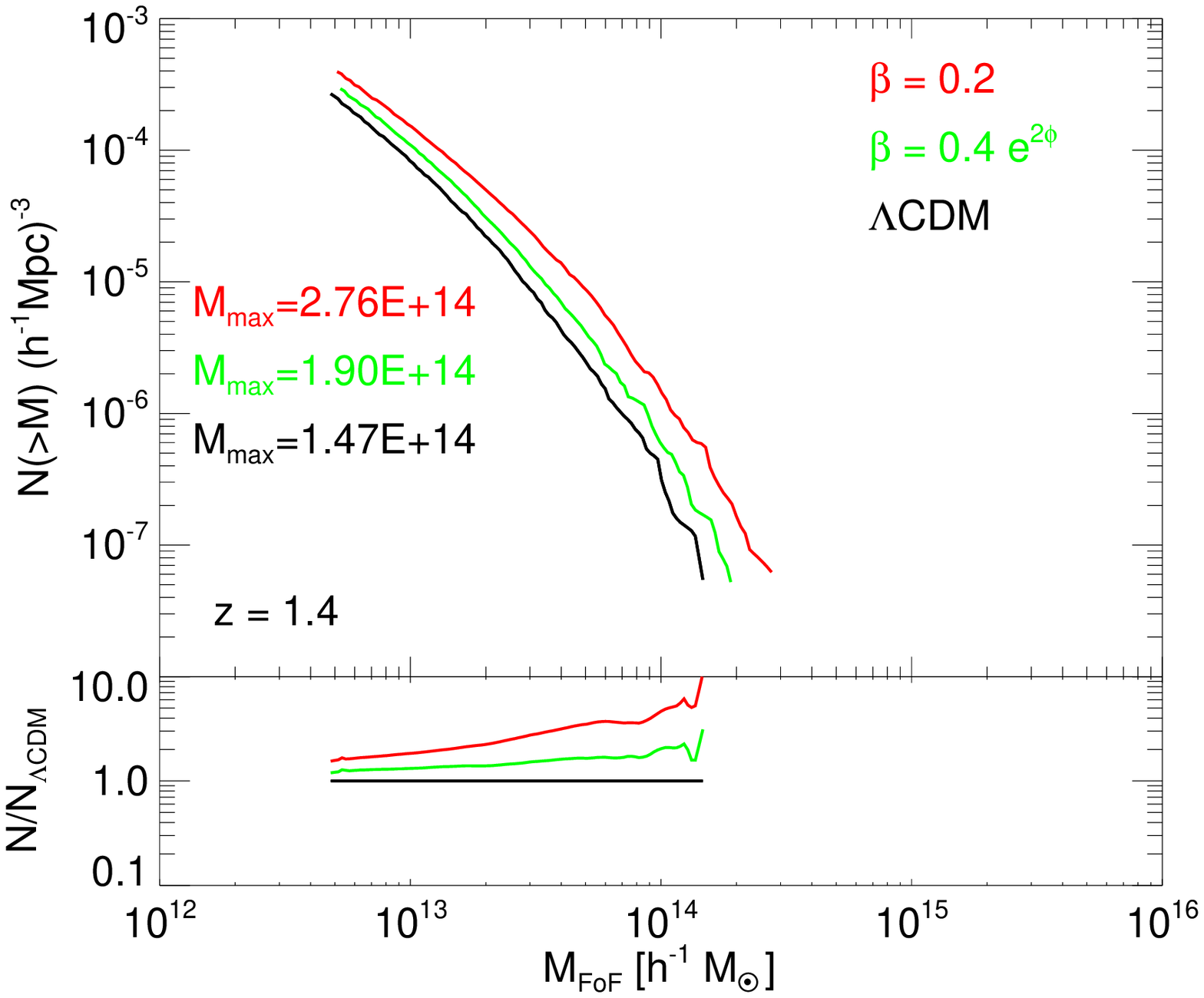} \\
\includegraphics[scale=0.45]{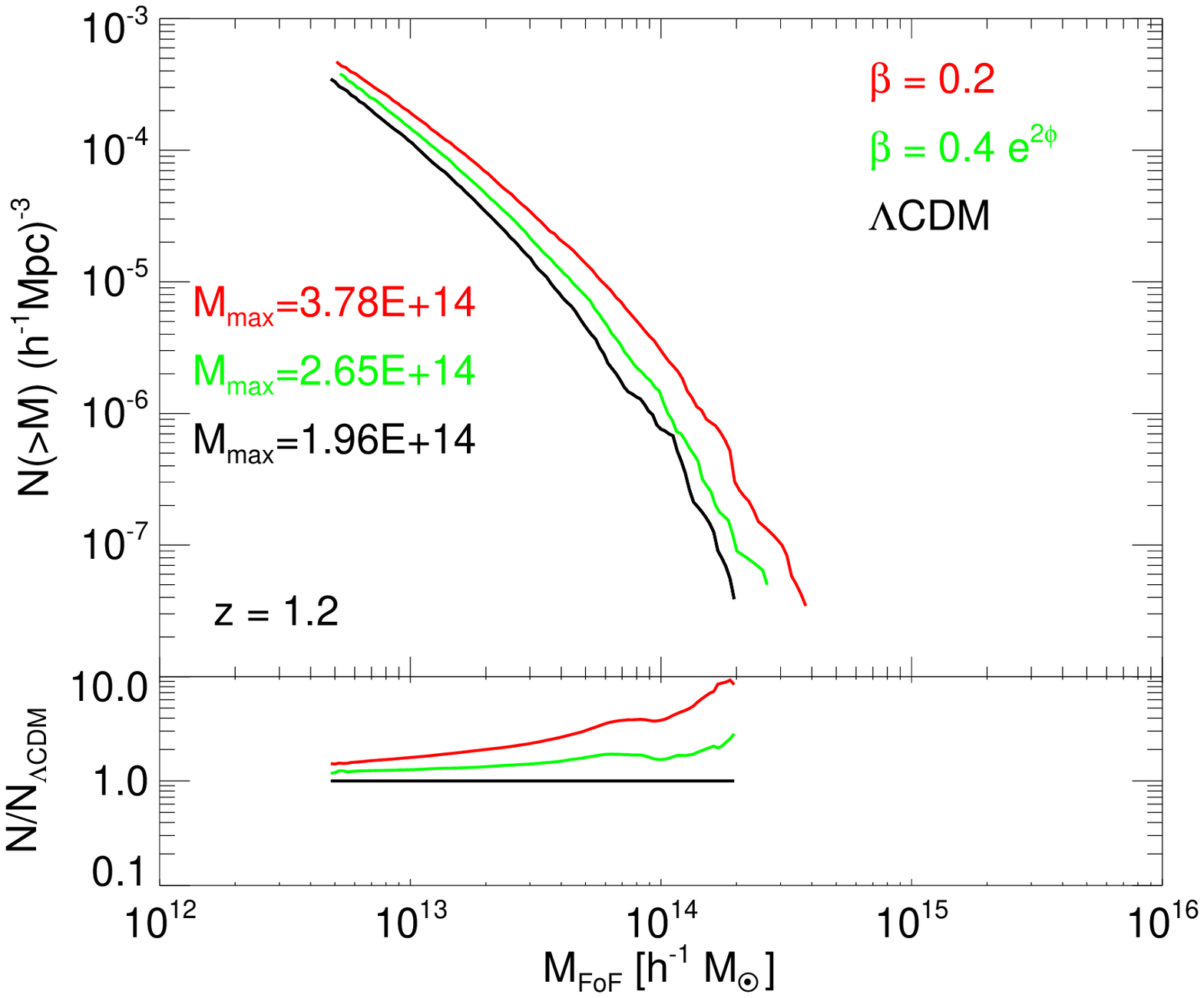}
\includegraphics[scale=0.45]{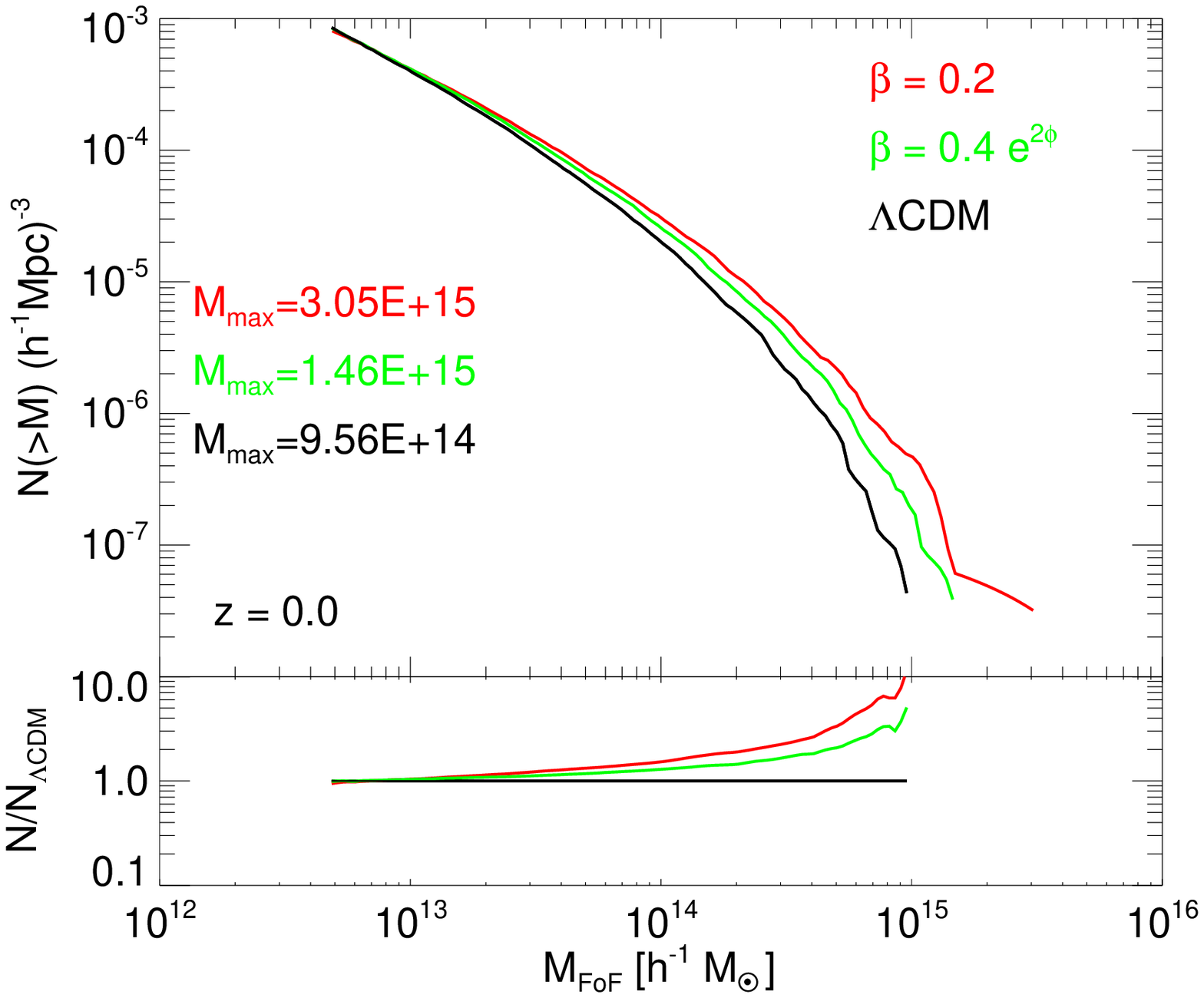}
  \caption{Cumulative mass functions for the low-resolution simulations of the coupled DE models with constant (red) and variable (green) couplings, and for the standard $\Lambda $CDM cosmology (black). The four plots correspond to different redshifts and show in the bottom panel the ratio of the halo number density to the $\Lambda $CDM case. The mass functions are based on the groups identified with a FoF algorithm and the masses quoted in the plots are FoF masses. Each plot also reports the mass of the most massive halo found at a given redshift in each simulation, which clearly shows how more massive structures are expected to form at any cosmological epoch in coupled dark energy models as compared with $\Lambda $CDM.}
\label{cumulative_massfunctions}
\end{figure*}

\begin{figure}
\includegraphics[scale=0.4]{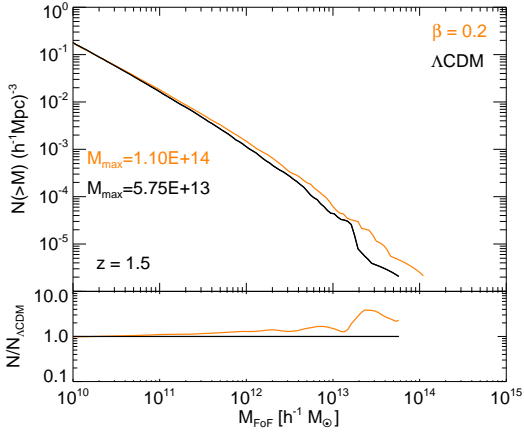}
  \caption{Cumulative mass function at $z =1.5$ for the coupled DE model RP of Table~\ref{Simulations_Table} (orange) compared to $\Lambda$CDM (black) from a high-resolution hydrodynamical simulation in a box of $80~h^{-1}$ comoving Mpc. The enhancement of the mass function in this set of simulations is slightly lower than in pure CDM simulations due to the presence of a fraction of uncoupled baryons. The bottom panel shows the enhancement of halo number density relative to $\Lambda$CDM.}
\label{mass_corr_plot}
\end{figure}

In Fig.~(\ref{cumulative_massfunctions}) we show the evolution of the halo mass function at different redshifts in the three low resolution simulations described in Table~\ref{Simulations_Table}. The bottom panel of each plot shows the enhancement of halo number density in coupled dark energy models with respect to $\Lambda $CDM. As it clearly appears in all the plots, the number density of halos of any mass is larger in coupled dark energy models, both for constant and variable couplings, as compared to $\Lambda $CDM, and the effect significantly grows with mass (the decreasing behavior found in \cite{Manera_Mota_2006} can be explained in view of \cite{Wintergerst_pettorino_2010}). In particular, at $z\approx 1.5$ the halo number density at the high-mass end of our simulated mass functions exceeds the $\Lambda $CDM value by a factor $\sim 10$ and $\sim 3$ for constant and variable couplings, respectively.

In Fig.~(\ref{mass_corr_plot}) we plot the same quantities for our hydrodynamical high-resolution simulations, which confirm the trend found in the larger simulation box, although the enhancement is slightly weaker in this case due to the inclusion of the fraction of uncoupled baryons.

For all simulations, the most massive halo forming in the simulated box has a higher mass for coupled dark energy cosmologies than for $\Lambda $CDM. In this respect, it is important to stress once more here that all the simulations start from the very same initial conditions, and therefore this effect is due only to the different physics induced by the coupling and is not affected by statistical differences in the random realization of the initial density fields. Due to our normalization choice, the different models will also be characterized by different values of $\sigma _{8}$ at $z=0$. In order to quantify this effect we have computed $\sigma _{8}(z=0)$ from our two hydrodynamical simulations. The maximum increase in $\sigma _{8}$ is obtained, as expected, for the highest value of the coupling: $\sigma _{8}(\beta = 0.2) \sim 0.91$. 
Compared to the $\Lambda $CDM value $\sigma_ {8}( \Lambda {\rm CDM}) \sim 0.76$, this corresponds to a maximum relative increase of roughly 20\%, still in reasonable agreement with present available measurements of $\sigma _{8}$ at $z=0$ \citep[see e.g.][]{Mantz_etal_2009,Vikhlinin_etal_2009}. We stress again  that different values of $\sigma _{8}$ at the present time arise from the same normalization at high redshift. In other words, different couplings are able to evolve the same initial normalization into different $\sigma _{8}$ today. As a consequence, a mismatch between the value of the perturbations amplitude at CMB and local measurements of $\sigma _{8}$ based solely on low redshift probes, could provide a further possible observational test for coupled dark energy models, breaking degeneracies with $\Lambda $CDM cosmologies. Interestingly, some mismatch between the value of $\sigma _{8}$ extrapolated from CMB data under the assumption of a standard $\Lambda $CDM cosmology and pure local measurements has been recently claimed by \eg \citet{Feldman_etal_2003,Reichardt_etal_2009,Watkins_etal_2009}, although still with a low statistical significance.

We note also that the effect is expected to be model-dependent and possibly weaker for other viable models as e.g. variable coupling scenarios.

\section{Conclusions}

We conclude that coupled dark energy can significantly increase the probability to detect massive clusters at high redshift with respect to a $\Lambda$CDM model. The effect shows a quite strong dependence on the halo mass and is found to be larger for higher halo masses.
Note that similar coupled dark energy models may also predict the existence of structures at very large scales \citep{Wintergerst_etal_2010}. 

For the case of a constant coupling we have assumed the largest possible value of the coupling which is allowed according to present constraints
for this class of models \citep{bean_etal_2008,lavacca_etal_2009,Baldi_Viel_2010}. In this case the number density of halos at $z\sim 1.5$ is found to be at least a factor $\sim 10$ higher than 
in $\Lambda$CDM cosmologies for halos of masses larger than $M\approx 10^{14}$ M$_{\odot }$. This is a conservative estimate and the enhancement could increase in case the models were normalized further in the past (\eg at decoupling) rather than at $z \sim 60$ as considered in the present simulations. 

We have also presented results for one choice of variable coupling models. Despite the large value of the interaction strength at recent cosmological
 epochs, such models have a weaker impact on the background expansion history as compared to constant coupling models and are therefore easier to reconcile with present cosmological constraints. For the specific choice of variable coupling that we have considered in this work, which has recently been shown
 to also mitigate other possible tensions between the $\Lambda $CDM cosmology and observations at small scales \citep[as \eg the
 cluster baryon fraction or the so called ``cusp-core" problem, see][]{Baldi_2010}, the total effect on the number counts 
 is found to be somewhat weaker than in the constant coupling scenario, 
 due to the low value of the coupling at early times. Nevertheless, also in this model the probability of finding massive clusters
 at $z\sim 1.4$ is enhanced by at least a factor $2-3$ with respect to $\Lambda$CDM.
 
These estimates are based on the ratio of the simulated halo mass functions at their high-mass end in the different models. 
Although the most massive groups found in our simulations at $z\sim 1.4$ do not reach the mass of the high-z cluster XMMU J2235.3-2557
 identified by \citet{Jee_etal_2009} and \citet{Rosati_etal_2009}, the extrapolation to slightly higher masses of the enhancement found
 in the number density at the high-mass end of our halo catalogs is a conservative approach since the effect is found
 to be quite strongly increasing with mass (see Figs.~\ref{cumulative_massfunctions},\ref{mass_corr_plot}). 
It is also interesting to notice that the mass of the most massive halo forming in coupled dark energy cosmologies, both for constant and variable couplings, is systematically larger than for the corresponding $\Lambda $CDM cosmology. The related increase of $\sigma _{8}$ does not exceed 20\% for the largest possible coupling and could be significantly reduced for other viable scenarios or by performing an optimization of the full set of cosmological parameters. In any case, such an increase concerns only the value of $\sigma _{8}$ at low redshifts, while all the models are consistent at high redshift with the present CMB data. One may want to compare the effects found in the present analysis for coupled dark energy with those of a $\Lambda$CDM model with a higher $\sigma _{8}$: we have noted that such a degeneracy may be broken by directly measuring the redshift evolution of $\sigma _{8}$, thereby providing an additional observational test for coupled dark energy as compared to $\Lambda $CDM.\\

We have therefore shown that both constant and variable coupling models, also for different scalar potentials, enhance at any cosmological epoch the cumulative halo mass function -- and consequently the probability to detect halos of any given mass in volume limited surveys -- with respect to the standard $\Lambda$CDM cosmology.
Future detection of massive clusters at high redshift can therefore be used to disentangle a cosmological constant
 from dynamical coupled dark energy models.

\section*{Acknowledgments.}
We thank L.Amendola and J.Norena for useful discussion. 
MB acknowledges SISSA for hosting him during the development of the present work.
MB is supported by the DFG Cluster of Excellence ``Origin and
  Structure of the Universe" and partly supported by the TRR
  Transregio Collaborative Research Network on the ``Dark
  Universe"

  \bibliographystyle{mn2e} \bibliography{baldi_bibliography.bib}

\end{document}